\documentclass[a4paper,10pt]{article}
\usepackage[utf8]{inputenc}
\usepackage{authblk}

\usepackage{listings}
\usepackage{graphicx}
\usepackage{placeins}
\usepackage{epstopdf}
\usepackage{todonotes}
\usepackage{lineno,hyperref}
\usepackage{upgreek}
\usepackage{units}
\usepackage{multirow}
\usepackage{array}
\usepackage{multicol}
\usepackage{amssymb}	
\usepackage{amsmath}	
\usepackage{textcomp}
\usepackage{pdfpages}

\usepackage{caption}
\usepackage{subcaption}

\title{Investigating the impact of the diode edge geometry on the effective active area of silicon sensors using AREA-X}

\author[1,2]{Luise Poley}
\author[3]{Ian Dyckes}
\author[1,2,4]{Javier Fern\'{a}ndez-Tejero}
\author[4]{Celeste Fleta}
\author[5]{Dennis Sperlich}
\author[4]{Miguel Ull\'{a}n}

\affil[1]{Department of Physics, Simon Fraser University, 8888 University Dr, Burnaby, V5A 1S6, BC, Canada}
\affil[2]{TRIUMF, 4004 Wesbrook Mall, Vancouver, V6T 2A3, BC, Canada}
\affil[3]{Lawrence Berkeley National Laboratory, Cyclotron Road, Berkeley, USA}
\affil[4]{Instituto de Microelectr\'{o}nica de Barcelona, (IMB-CNM, CSIC), Campus UAB-Bellaterra, Carrer dels Tillers, 08193, Barcelona, Spain}
\affil[5]{Physikalisches Institut, Albert-Ludwigs-Universit\"{a}t Freiburg, Hermann-Herder-Stra\ss{}e, Freiburg im Breisgau, Germany}

\begin{document}

\maketitle

\begin{abstract}

Sensors for particle tracking detectors are required to provide a maximum active area in addition to fulfilling performance criteria concerning radiation hardness, charge collection and operating conditions (e.g. leakage current, depletion voltage and breakdown voltage). While the requirement for optimised coverage within the tracking detector necessitates a slim sensor edge between active region and physical sensor edge, wider edge regions were found to be beneficial for the sensor performance during an early prototyping phase. In order to study the impact of differently sized edge regions, test structures were used to compare their individual active regions.

Measurements of each diode were performed using a micro-focused X-ray beam to map its respective active area. This paper presents measurements of these test structures using AREA-X showing that the active area of a silicon particle tracking sensor does not only depend on the size of its bias ring, but also the size and configuration of its edge structure.

\end{abstract}

\section{Introduction} 

Sensors for particle tracking detectors are optimised to maximise their active area, i.e. the area over which traversing particles can be detected, compared to their inactive area. For typical silicon sensors, a maximum active area is achieved by minimising the distance between the bias ring, which surrounds the active area, and the edge ring, which defines the outer perimeter next to the dicing edge.

As part of the development of silicon strip sensors for the ATLAS Inner Tracker (ITk), the investigation of different edge geometries indicated~\cite{edge} that a closer proximity between bias ring and physical sensor edge increased the chances for the occurrence of a premature breakdown (where the sensor breakdown voltage is lower than the required operating voltage). In order to quantify this connection, diodes with varying distances between bias ring and edge ring were designed as part of the test structures of a wafer prototype for the ATLAS ITk.

While a larger distance between bias ring and edge ring was expected to avoid premature breakdown, it was also assumed to increase the inactive region of the sensor, i.e. affect the ratio between active and inactive sensor regions adversely.

Therefore, measurements were performed on an array of diodes with identical implant areas and varying edge regions to study their impact on the width of the corresponding active diode region.

\section{Device under Test}

\begin{figure}[]
\centering
\includegraphics[width=1.0\linewidth]{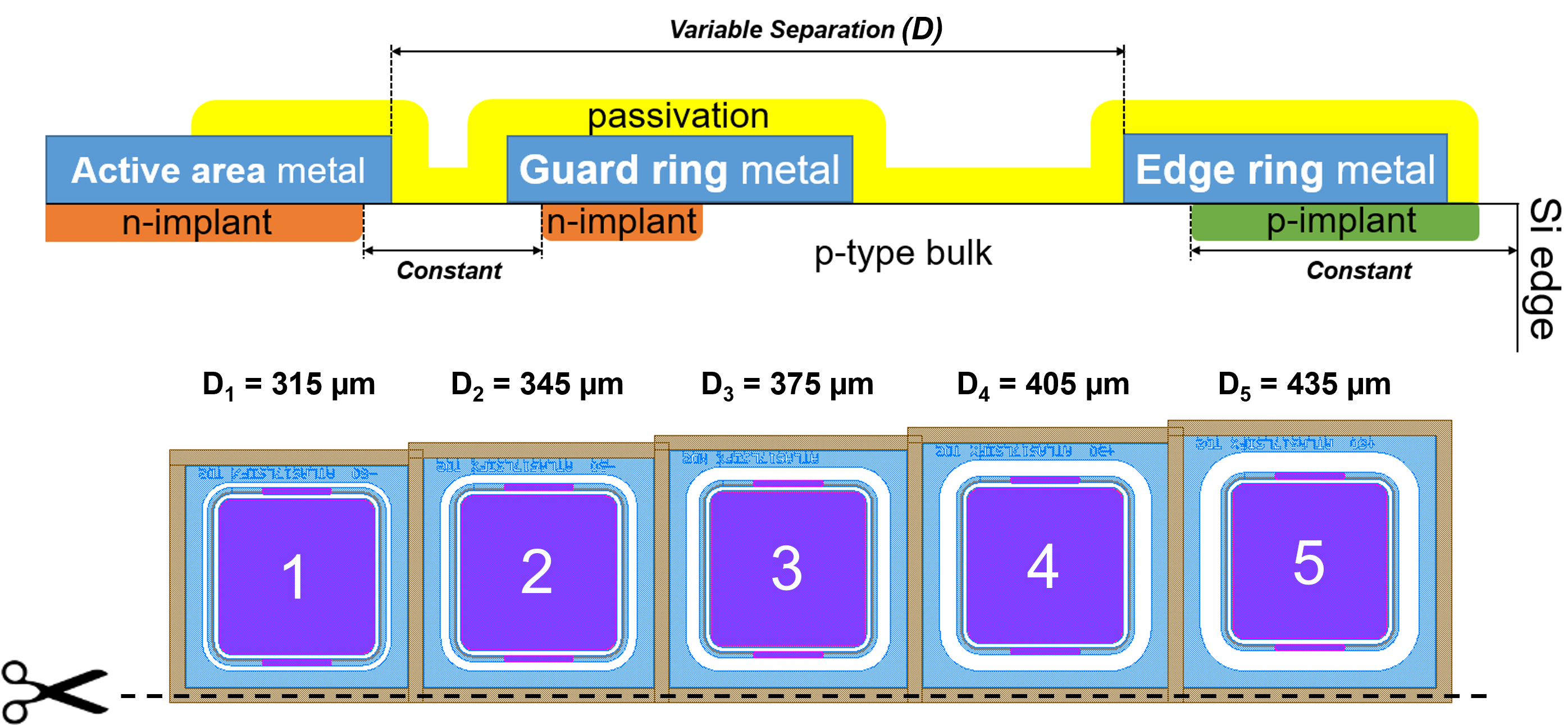}
\caption{Test structure designed to study the influence of the sensor physical edge, respect to the active area. Five {\unit[2$\times$2]{mm$^{2}$}} diodes with variations on the edge distance.}
\label{fig:layout}
\end{figure}

The test structure under investigation was added to the 6-inch wafer layout of the prototype ATLAS17LS\cite{ATLAS17LS-IFX} for the participation of Infineon Technologies A.G. in the Market Survey for the ATLAS ITk strip sensors~\cite{Xavi_thesis}. The test structure is composed of an array of five \unit[2$\times$2]{mm$^{2}$} diodes with identical implant area and guard ring, and one of the dicing lines aligned. The central diode has an edge configuration identical to the ATLAS17LS full-size sensor, and the rest of the diodes have variable distances between implant area and silicon physical edge, in steps of \unit[30]{$\upmu$m}, resulting in edge distances going from 315 to \unit[435]{$\upmu$m} (see figure~\ref{fig:layout})

The array of diodes was tested as one complete structure, diced from an ATLAS17LS wafer with a p-doped bulk and n-doped implants. They have a thickness of \unit[300]{$\upmu$m} and a depletion voltage of approximately \unit[-300]{V}~\cite{ATLAS17LS-IFX}.

\begin{figure}[]
\centering
\includegraphics[width=1.0\linewidth]{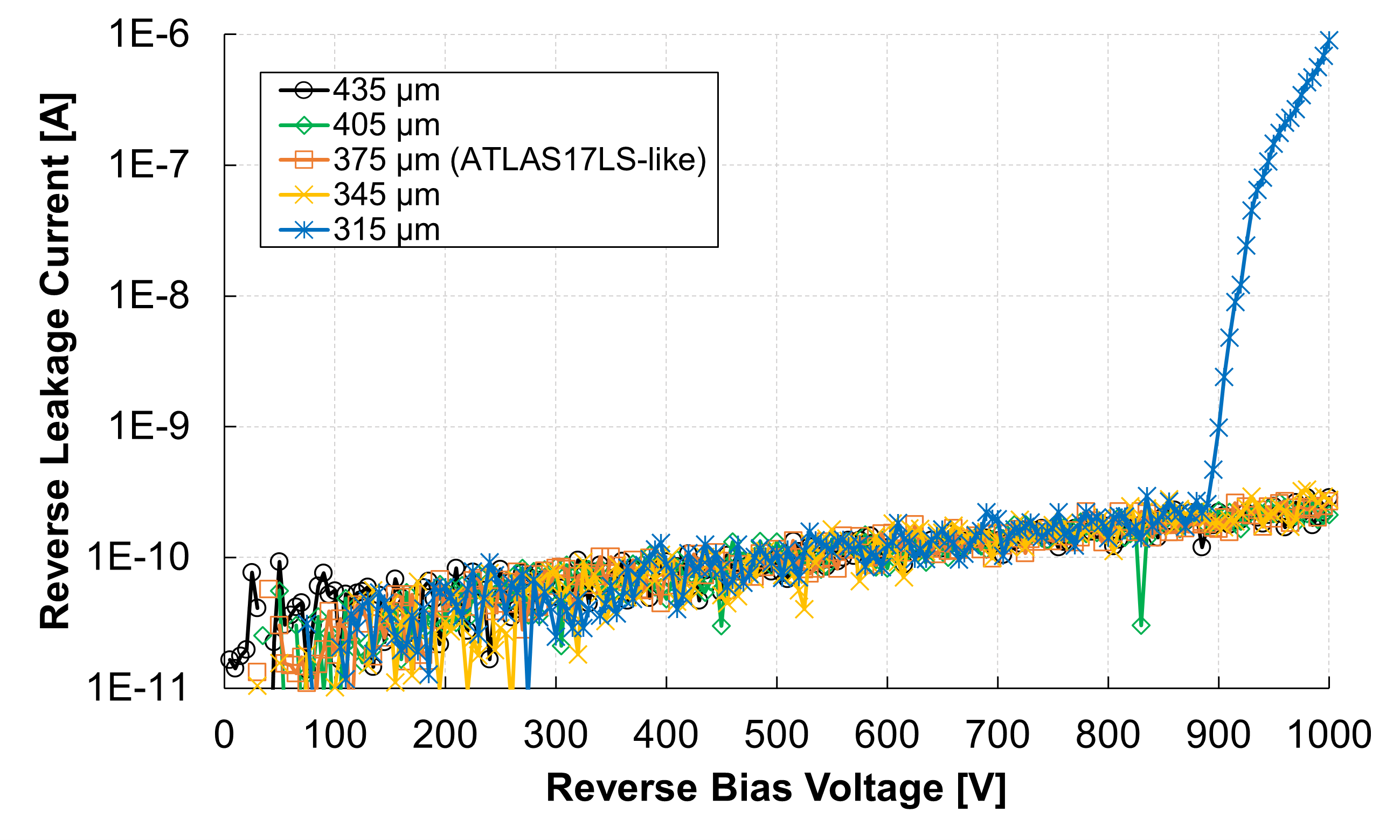}
\caption{Leakage current measured for the five diodes with variations on the edge distance.}
\label{fig:IVs}
\end{figure}

In a previous study\cite{Xavi_thesis}, the leakage current of the different diodes was measured, showing similar leakage current, and no breakdown below \unit[-1]{kV}, except the diode with narrower edge (\unit[315]{$\upmu$m}) showing a breakdown at \unit[-900]{V} (figure~\ref{fig:IVs}).

\section{Measurement}

The diode array was tested using AREA-X\cite{AREAX} utilising a monochromatic, micro-focused X-ray beam with a diameter of \unit[$2\times3$]{$\upmu$m} and an energy of \unit[15]{keV} provided by beamline B16 at the Diamond Light Source~\cite{B16}. The beam was pointed at the diode array, biased to \unit[-400]{V} and therefore assumed to be fully depleted.

A \unit[15]{keV} photon traversing a sensor volume with a thickness of \unit[300]{$\upmu$m} has a \unit[51]{\%} probability to interact within the sensor, which leads to the transfer of its energy to a single electron, which travels between \unit[10-20]{$\upmu$m} within silicon before its energy is absorbed. Each individual photon thereby produces about 4,200 electron-hole-pairs, which result in an induced photo current which adds to the overall sensor leakage current if the electron is traveling through a volume of the sensor that has been depleted. By monitoring the leakage current while moving the beam across the diode under test, the resulting photo current can be used to map the depleted area of the device.

The measurements were performed by mounting the diode array on a printed circuit board designed for simultaneous readout of sensor current and temperature, mounted inside a closed dark box providing cooling down to \unit[-7]{$^{\circ}$C}. The dark box was mounted on precision xy-stages and moved in steps of \unit[50]{$\upmu$m} across the full area of the structure under test. After each position change, the sensor current was read out and stored after a \unit[6]{s} settling time to prevent current fluctuations. The stage was moved across the shorter direction of the diode array before moving to the next position along the longer edge of the structure.

Throughout the measurement, the following parameters were monitored
\begin{itemize}
 \item device temperature
 \item humidity inside the dark box
 \item beam current, i.e. photon flux
\end{itemize}
using both monitoring provided by the beamline and instrumented in the measurement setup.

\section{Results}

\begin{figure}
    \begin{subfigure}{\textwidth}
    \includegraphics[width=\linewidth]{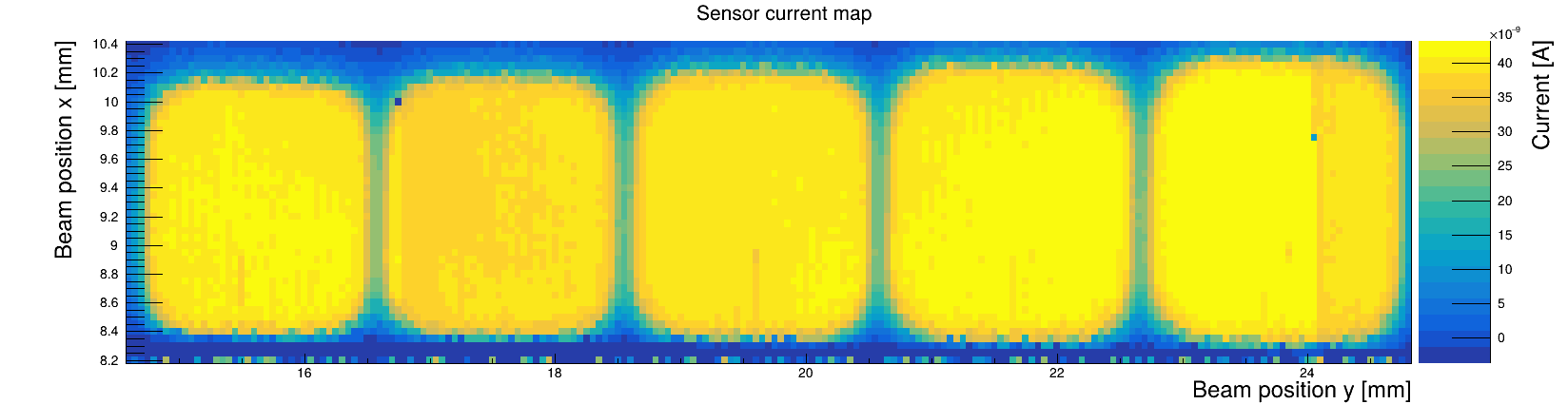}
    \caption{Sensor current as measured}
    \label{fig:map_1}
    \end{subfigure}
    \begin{subfigure}{\textwidth}
    \includegraphics[width=\linewidth]{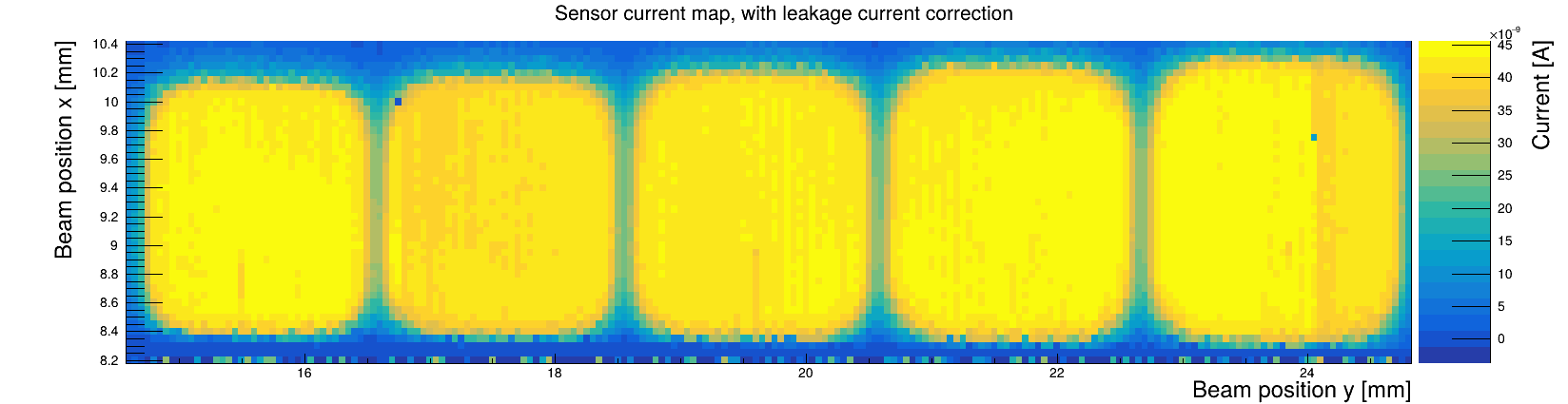}
    \caption{Sensor current, corrected for leakage current}
    \label{fig:map_3}
    \end{subfigure}
    \begin{subfigure}{\textwidth}
    \includegraphics[width=\linewidth]{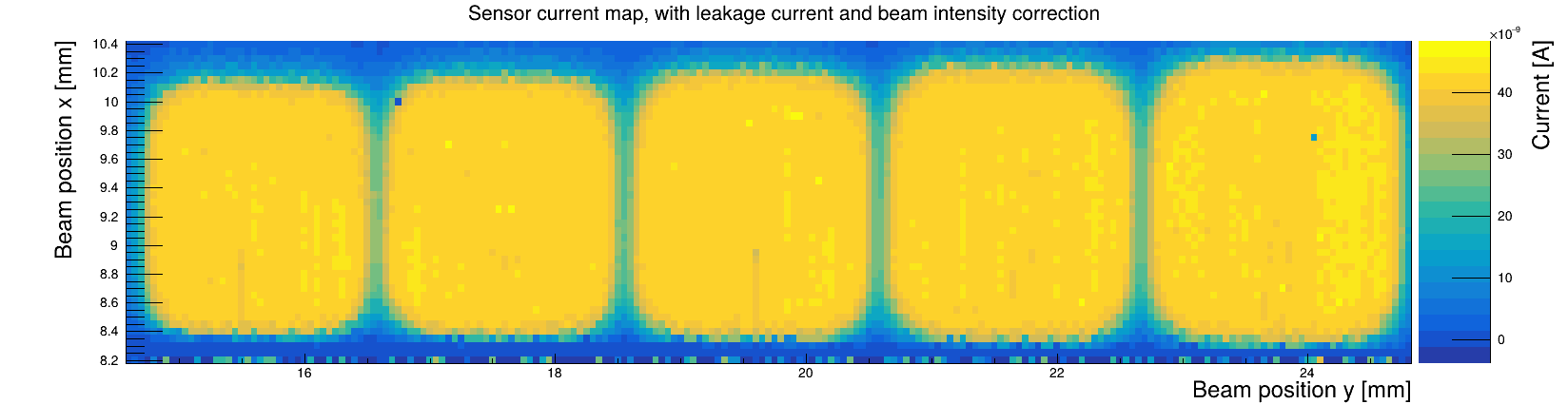}
    \caption{Sensor current, corrected for leakage current and beam intensity}
    \label{fig:map_4}
    \end{subfigure}
    \begin{subfigure}{\textwidth}
    \includegraphics[width=\linewidth]{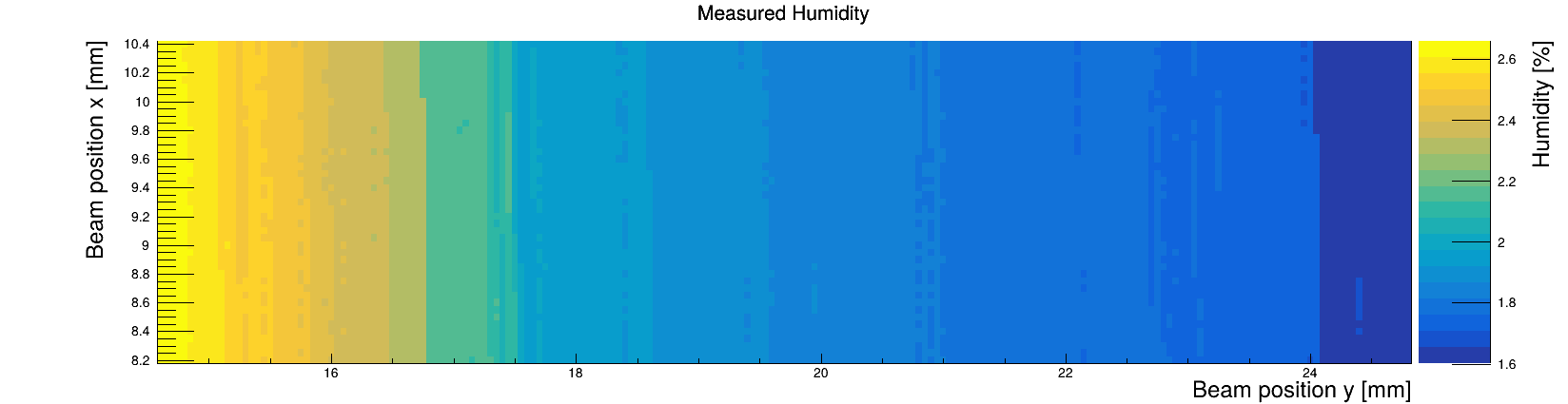}
    \caption{Humidity measured over the course of the beam test - over the course of the measurement, the humidity had dropped from {\unit[2.6]{\%}} to {\unit[1.6]{\%}}.}
    \label{fig:map_hum}
    \end{subfigure}
\caption{Sensor current measured for the diode array under test, corrected for known systematic effects. In the orientation of the displayed maps, the stage was moved vertically with respect to the beam to complete each column before moving to the next one.}
\end{figure}

The sensor current obtained for each position of the beam with respect to the device is shown in figure~\ref{fig:map_1}. 

Comparing the current observed in depleted areas of the sensor ($\mathcal{O}$(\unit[10$^{-8}$]{A}), corresponding to leakage current and photo current) and outside the depleted volume ($\mathcal{O}$(\unit[10$^{-10}$]{A})leakage current), the sensor current was found to be dominated by the photo current. Therefore, no temperature correction was applied to the measured sensor current, as it should only be applied to the leakage current, not the photo current.

Instead, the overall leakage current was subtracted from the total sensor current using the current measured when the beam was pointing to areas outside the depleted sensor volume (see figure~\ref{fig:map_3}).

Finally, the beam current, measured using a diode in the beam path, was used to scale the remaining photo current (see figure~\ref{fig:map_4}). Subsequent analyses were performed using the map obtained from applying both leakage current and beam current corrections.

In order to extract effective active areas of each diode in the array, they were individually analysed per column of the beam positions.

\paragraph{1. Determining the central plateau height}

\begin{figure}[]
\centering
\includegraphics[width=1.0\linewidth]{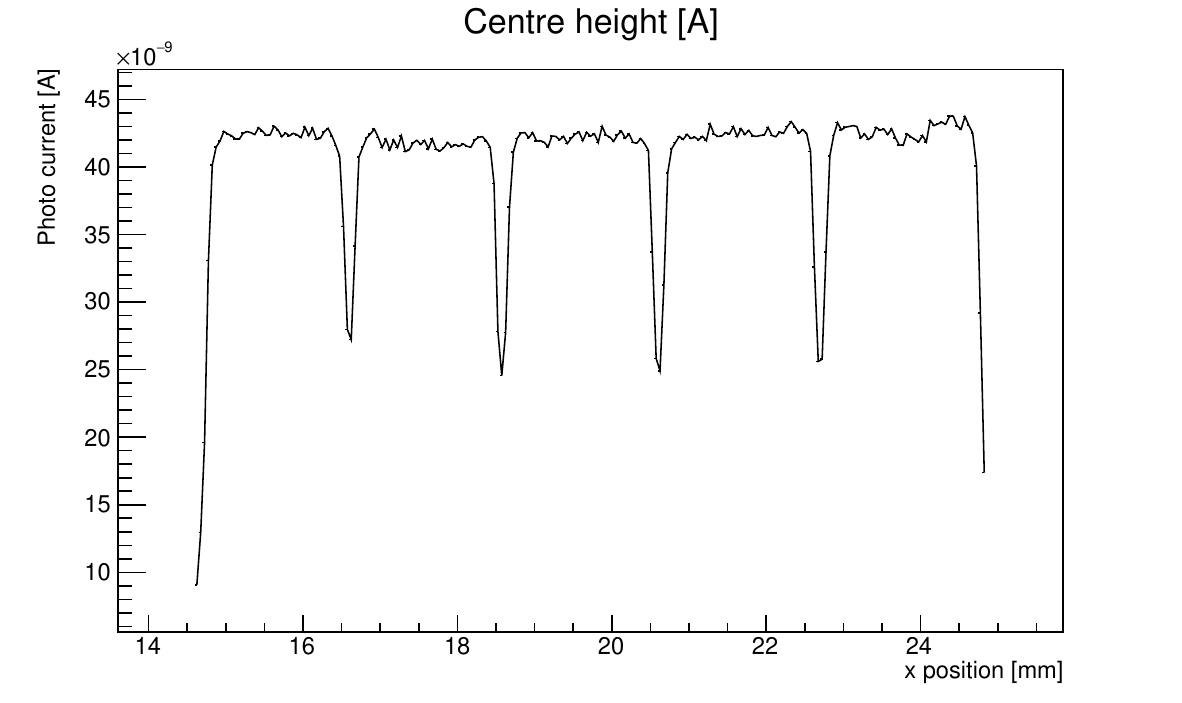}
\caption{Plateau heights determined per column of beam positions in X. The determined heights were found to be consistent across the diode array.}
\label{fig:centre_height}
\end{figure}

For each column of beam positions i.e. for each X position of the beam, the height (photo current) of the central plateau was determined from a fit at the centre of the diode (see figure~\ref{fig:centre_height}).
The obtained plateau heights per column (average centre photo currents) were then used for the subsequent analysis steps.

\paragraph{2. Determining the edges of the diode}

\begin{figure}[]
\centering
\includegraphics[width=1.0\linewidth]{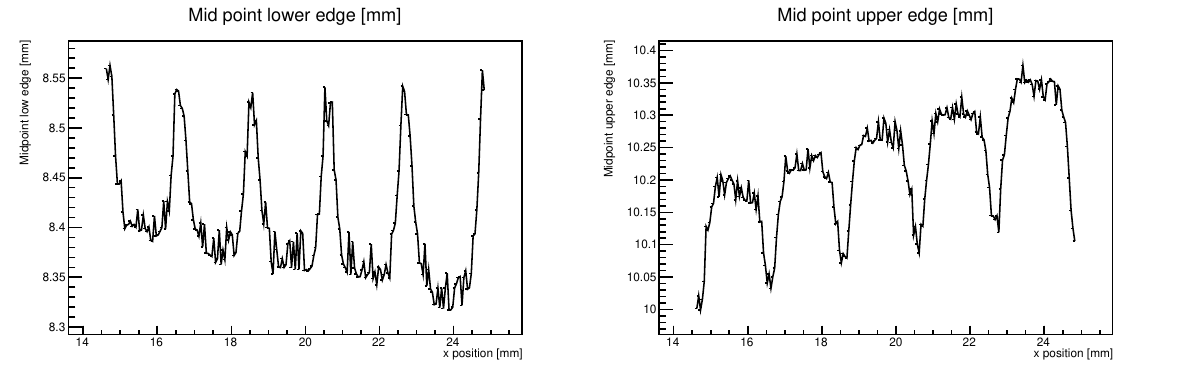}
\caption{Upper and lower edge for each column of beam position determined from the point at which the leakage current and beam intensity corrected photo current had dropped to {\unit[50]{\%}} of the plateau.}
\label{fig:edges}
\end{figure}

Using the plateau height determined in the previous analysis step, the active width of each column was then determined by identifying the width between the points at \unit[50]{\%} of the maximum (called "edges") (see figure~\ref{fig:edges}). 
Across the central region of each diode, the position of each edge was found to be consistent within about \unit[5]{$\upmu$m}. Due to corner effects, the edges of the active region were found to move closer towards the centre at the outer diode regions.

\paragraph{3. Determining the full diode width}

\begin{figure}[]
\centering
\includegraphics[width=1.0\linewidth]{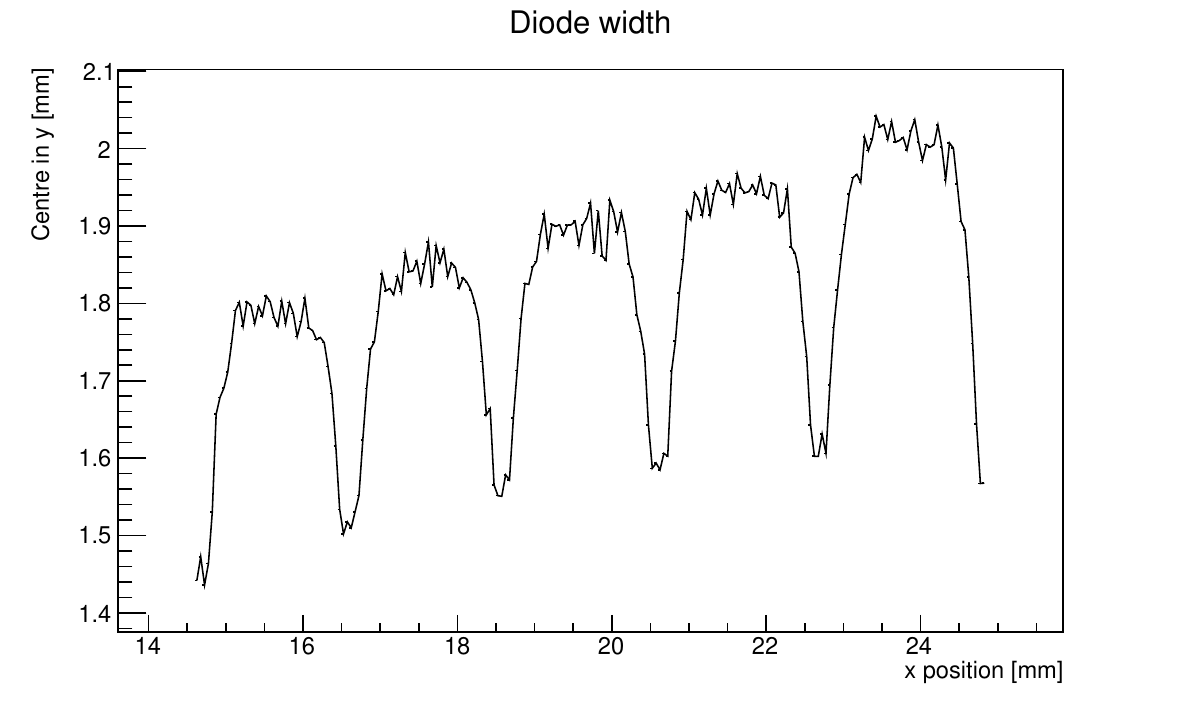}
\caption{Width of the diode for each column of beam positions: despite identical inner designs, the diodes show an increasing width.}
\label{fig:width}
\end{figure}

The full width of the active area was determined from the distance between the upper and lower edge per column (see figure~\ref{fig:width}).

The active width per column was then used to calculate the average active width for each diode (see table~\ref{tab:width}).
\begin{center}
\begin{table}
\begin{tabular}{l|rrrrr}
 & \multicolumn{5}{c}{Diode}\\
 & 1 & 2 & 3 & 4 & 5\\
\hline
Active width $[$mm$]$ & 1.78 & 1.84 & 1.89 & 1.94 & 2.01 \\
Width inside bias ring $[$mm$]$ & 1.25 & 1.25 & 1.25 & 1.25 & 1.25 \\
Width between edge implants $[$mm$]$ & 1.56 & 1.62 & 1.68 & 1.74 & 1.80 \\
\end{tabular}
\caption{Calculated width of each diode in comparison with its geometrical size}
\label{tab:width}
\end{table}
\end{center}

\begin{figure}
    \begin{subfigure}{\textwidth}
    \includegraphics[width=\linewidth]{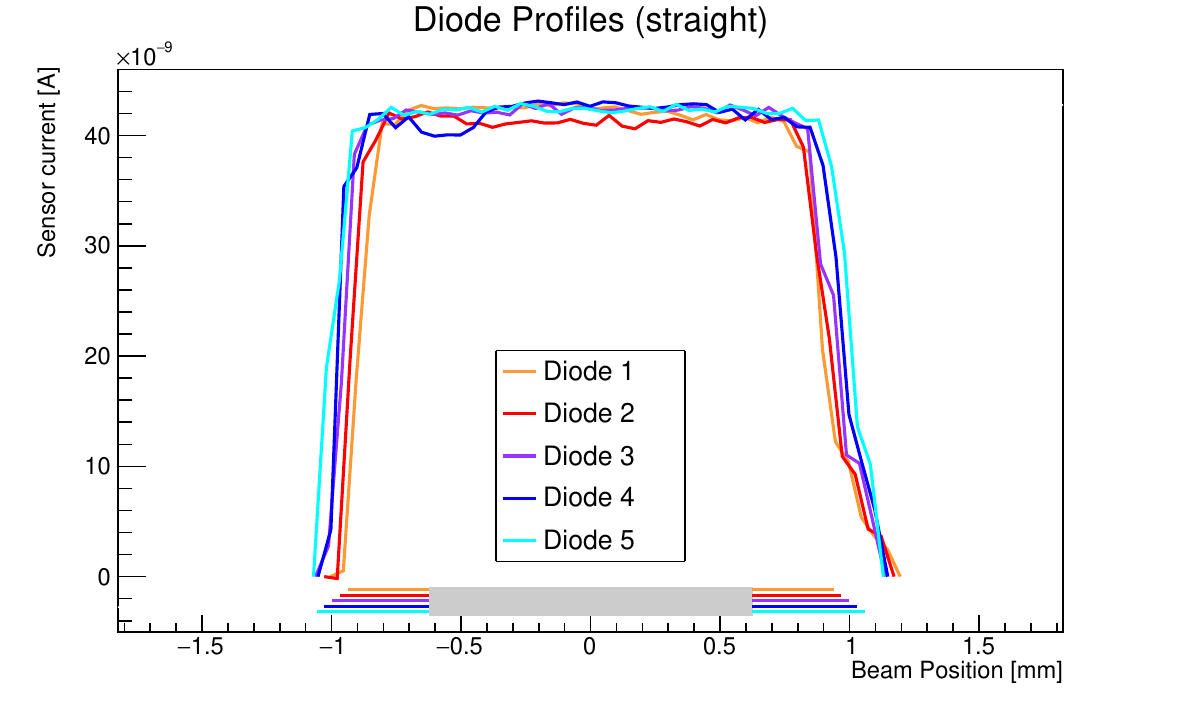}
    \caption{Response across diodes (straight) in comparison with the implant width (grey) and the distance to edge ring for each diode (indicated in the same colour as the current response).}
    \label{fig:straights}
    \end{subfigure}
    \begin{subfigure}{\textwidth}
    \includegraphics[width=\linewidth]{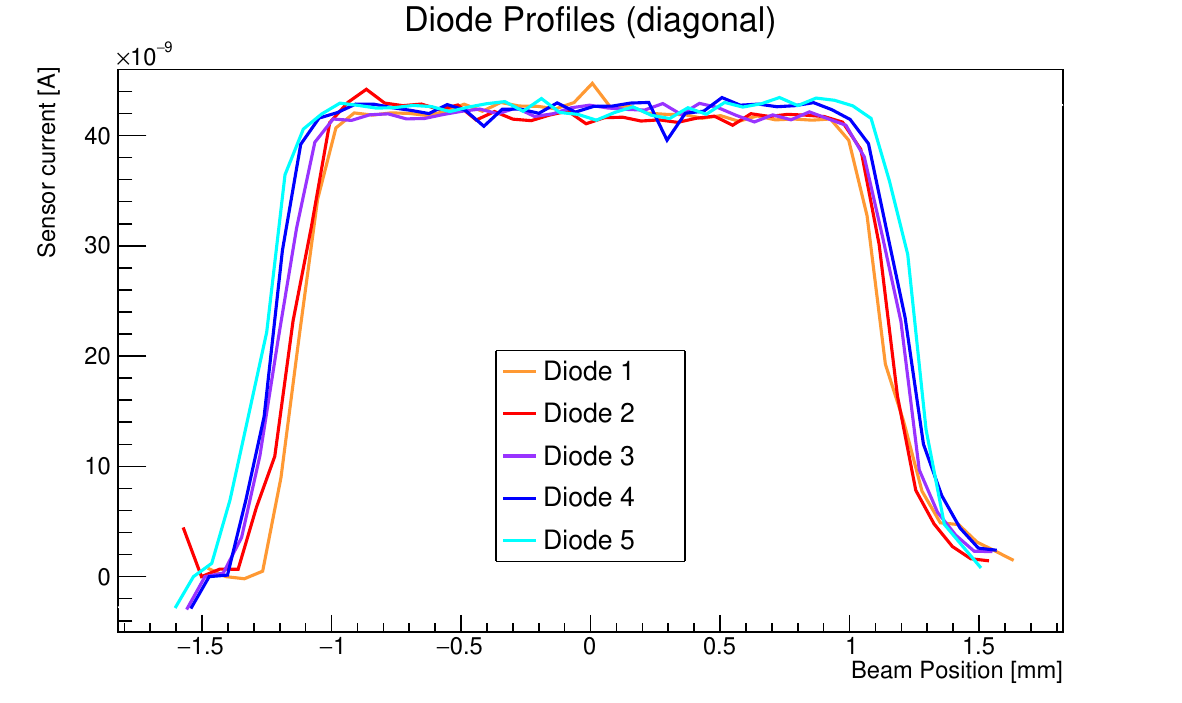}
    \caption{Response across diodes (diagonal)}
    \label{fig:diagonals}
    \end{subfigure}
\caption{Responses across diodes in straight and diagonal direction: in both cases, the width of the active area increases with the size of the edge ring despite equal sizes of bias and guard rings}
\end{figure}

The measurement showed that the active width of the diode increased with the size of the edge ring rather than with the size of the bias or the guard ring. A direct comparison of the extent of all diodes in comparison can be found in figures~\ref{fig:straights} and~\ref{fig:diagonals}.
Similar to the active width measured across the individual diodes, the active widths in diagonal direction were found to agree well with the size of their edge structures.

\FloatBarrier

\section{Conclusion and Outlook}

Measurements were conducted on an array of diodes designed to study the impact of the diode geometry on their resulting active area using a micro-focused X-ray beam. The obtained results showed that the size of the active area is directly correlated with the size of the edge ring instead of those of the the implant or guard ring.

Another series of measurements is planned in order to investigate the same correlation for under-depleted and irradiated diodes. 

\section*{Acknowledgements}

The authors would like to thank the Diamond Light Source for access to beam line B16 (proposal OM32397), where this method was developed and tested, especially Oliver Fox and Kawal Sawhney, for providing advice, support and maintenance during the experiment.

This work was in part supported by the Canada Foundation for Innovation and the Natural Sciences and Engineering Research Council of Canada as well as the Alexander von Humboldt Foundation. This work is supported and financed in part of the Spanish R\&D grant PID2021-126327OB-C22, funded by MICIU/AEI/ 10.13039/501100011033 and by ERDF/EU.

\bibliographystyle{unsrt}
\bibliography{bibliography.bib}

\end{document}